\documentclass[%
twocolumn,
 amsmath,amssymb,
 aps,
pra,
]{revtex4-2}
\usepackage{graphicx}% Include figure files
\usepackage{dcolumn}% Align table columns on decimal point
\usepackage{bm}% bold math
\usepackage{multirow,amsthm,amssymb,amsbsy,amsmath,epsfig,epstopdf,bm,float}
\usepackage{hyperref}
\usepackage{booktabs}
\usepackage{verbatim}
\usepackage{dcolumn}
\usepackage{color}
\usepackage[dvipsnames]{xcolor}
\usepackage[normalem]{ulem}
\usepackage{soul}
\usepackage{braket}
\usepackage[export]{adjustbox}

\newcommand{\be}{\begin{equation}}
\newcommand{\ee}{\end{equation}}

\newcommand{\bs}{\begin{split}}
\newcommand{\es}{\end{split}}

\begin{document}

%\preprint{APS/123-QED}

\title{Unveiling hidden features of the Kitaev model through a complex-network analysis }% Force line breaks with \\

\author{Guillem Llodr\`{a}}
\affiliation{Instituto de F\'{i}sica Interdisciplinar y Sistemas Complejos (IFISC, UIB-CSIC), Campus Universitat de les Illes Balears E-07122, Palma de Mallorca, Spain}
\author{Roberta Zambrini}%
 \email{roberta@ifisc.uib-csic.es}
\affiliation{Instituto de F\'{i}sica Interdisciplinar y Sistemas Complejos (IFISC, UIB-CSIC), Campus Universitat de les Illes Balears E-07122, Palma de Mallorca, Spain}

\author{Gian Luca Giorgi}%
\affiliation{Instituto de F\'{i}sica Interdisciplinar y Sistemas Complejos (IFISC, UIB-CSIC), Campus Universitat de les Illes Balears E-07122, Palma de Mallorca, Spain}


\date{\today}% It is always \today, today,

\begin{abstract}

We introduce a density matrix-based network analysis to explore the ground state of the Kitaev chain, uncovering previously hidden structural and entanglement features. This approach successfully identifies the critical point associated with the topological phase transition and reveals a singular point where the ground state exhibits uniform, nonzero entanglement between all fermion pairs, corresponding to a fully connected network structure. We provide an analytical explanation for this singular behavior and establish a connection to the concept of ground state factorization observed in spin chains. Moreover, we analyze the open chain scenario and observe characteristic symmetry changes in the ground state corresponding to Majorana zero modes. While complex network theory has been employed in the study of quantum systems before, we demonstrate that it can uncover fundamentally new physical insights. %\sout{These results demonstrate the power of complex network analysis in revealing hidden features in such systems.}

%In this work, we perform a density matrix-based network analysis of the Kitaev model to uncover its hidden properties. Our results demonstrate the effectiveness of this approach in capturing the critical topological point and reveal a previously unexplored phenomenon:  for specific parameter regimes, the ground state displays uniform non-zero entanglement across all fermion pairs. This behavior manifests as a complete graph in the network representation. We provide an analytical explanation for this intriguing phenomenon and make connections to the well-established concept of ground state factorization observed in spin chains. Moreover, we analyze the open chain scenario and observe characteristic symmetry changes in the ground state corresponding to Majorana zero modes. These results demonstrate the power of complex network analysis in revealing hidden features in such systems. %This work demonstrates the power of complex network analysis in revealing hidden features of quantum systems.

\end{abstract}

\maketitle

%\tableofcontents

%---------------------------------------------------------%
\section{Introduction}
%---------------------------------------------------------%

%In recent decades, network science has provided increasingly important tools for describing and understanding emergent properties of complex systems \cite{Barabasi16}. Indeed, networks are widely used in fields as diverse as artificial intelligence, social science, biology, engineering, and telecommunications.   In physics, for example, network theory has been used in statistical physics and condensed matter \cite{Bianconi01,Barabasi02,Dorogovtsev08}, and more recently in quantum information science and technology \cite{Biamonte2019}.

Over the past two decades, network science has emerged as a robust and versatile framework with applications in a diverse range of disciplines \cite{Barabasi16, Newman}. By analyzing the connectivity pattern between the nodes, network science has become a fundamental tool for investigating emergent behavior, phase transitions, and critical phenomena in areas such as statistical mechanics and condensed matter — areas where traditional models are often inadequate \cite{Bianconi01, Barabasi_critical, Dorogovtsev_phase, erdos1960evolution}. %, that traditional models do not always capture. 
Fundamental advances, such as the discovery of the small-world properties \cite{small_world}, scale-free network structures \cite{Barabasi_scale} and community detection methods \cite{community_detection} have significantly shaped the field, offering new insights into the organization of complex networks. Beyond these theoretical foundations, network science has found widespread applications in economics \cite{economics_1, economics_2}, sociology \cite{watts2004new, borgatti2009network, sociology_1}, and neuroscience \cite{neuroscience_2, neuroscience_3}, where analyzing connectivity patterns can predict system behaviors and unveil hidden structures. 

More recently, network theory met quantum information science, driving advancements in quantum communication, computation, and simulation \cite{Biamonte2019, bianconi2015interdisciplinary}. Quantum networks offer novel ways to process and transfer information. Notably, their ability to distribute entanglement over long distances has paved the way for the development of the quantum internet, a global network with the potential to transform information processing and communication \cite{kimble2008quantum}, with satellite links predicted to yield small-world topologies \cite{brito_satellite}. Two approaches have been extensively studied for constructing quantum networks; for further approaches, see \cite{nokkala2024complex}. The first approach relies on entanglement-based connections, where entangled states serve as the links between nodes \cite{duan2001long, wengerowsky2018entanglement}. %\cite{patil2022entanglement}
This architecture is particularly valuable for secure quantum communication \cite{ursin2007entanglement, scarani2009security, bedington2017progress}. The second approach involves physical connections between spatially separated quantum components \cite{hartmann2006strongly, le2016many}. This framework is especially relevant in quantum circuits, quantum annealing, quantum synchronization, and quantum reservoir computing, where the complex system dynamics enhance information processing \cite{arute2019quantum, simulators_carleo, kadowaki1998quantum,cabot2019quantum,fujii2017harnessing, mujal2021opportunities, llodra2024quantum}. The broader implications of quantum networks extend to other interdisciplinary fields, including quantum biology \cite{plenio2008dephasing,Lambert2013}, where network theory is used to model energy transfer in photosynthetic systems.

Lately, new approaches using complex network theory have been applied to quantum science. Ref. \cite{PhysRevX.14.021029} introduced wave-function networks to map quantum many-body features onto networks, with nodes as basis states and edges as distance metrics. Revealing network properties, such as scale-free behavior, helps to analyze the robustness of quantum simulator data. %Additionally, Kolmogorov complexity is used to quantify how complexity evolves across quantum phase transitions, highlighting new insights into critical phenomena.
%This finding demonstrates the potential of network-based descriptions to offer deeper insights into quantum complexity.

Another approach, introduced in Refs. \cite{carr2017,sundar2018complex} constructs networks based on correlations between quantum components. % In this framework, links in the network represent bipartite quantum correlations within a multipartite state. %This method was independently developed by \cite{chou2014network} and \cite{carr2017,sundar2018complex} to study the quantum phase transition in the 1D and 2D Kitaev Hamiltonians and an Ising spin, chain, respectively. 
This method, which makes use  of the correlations in the ground state's reduced density matrix, was applied to study quantum phase transitions in Ising spin chains. Independently, an early work \cite{chou2014network} developed a similar method to analyze topological transitions in 1D and 2D Kitaev Hamiltonians, but instead of using the reduced density matrix, it constructed the correlation network using the pairing amplitude over the ground state wave function.  The framework proposed in \cite{carr2017,sundar2018complex} was then used to detect quantum critical points in the Fermi-Hubbard model \cite{Bagrov2020} and the onset of Bose-Einstein condensation \cite{PhysRevE.100.012304}. Other applications include detecting entanglement structures \cite{sokolov2020emergent}, optimizing quantum state tomography \cite{garcia2020pairwise}, analyzing photon-number correlations \cite{Walschaers_2023}, studying decoherence effects \cite{Sundar2021}, assisting deep-learning models for quantum correlations \cite{koutny2023deep}, and understanding correlations in quantum cellular automata \cite{Jones2022}. This powerful approach extracts key information from quantum states without requiring full-density matrix reconstruction, which becomes infeasible for large systems.
Building on Refs. \cite{carr2017,sundar2018complex}, we use quantum correlation-based networks %from \cite{carr2017,sundar2018complex} 
and uncover a previously unnoticed feature of a quantum system. Applying this method to the finite-size Kitaev p-wave model \cite{Kitaev_2001}, we first detect the topological critical point through network analysis. 
%More strikingly, we \glb{find a} regime where the ground state forms a completely regular network with uniform connections and a global clustering degree of one, \gl{which corresponds to a homogeneous density matrix invariant under any site-site permutation}. 
More strikingly, we identify a parameter regime in which the ground state forms a fully regular network —characterized by uniform connections and a global clustering coefficient of one— corresponding to a homogeneous density matrix that remains invariant under arbitrary site permutations and exhibits uniform nonzero entanglement
between any pair of fermions.
We analytically link this phenomenon to ground state factorization in spin chains \cite{KURMANN1982235,Giampaolo}, highlighting the power of the complex network approach to unveil hidden features. Finally, in the open chain scenario, we track symmetry changes related to Majorana zero modes.

\section{The finite-size Kitaev model}\label{sec:model}   %
%----------------------------------------------------------%

The Kitaev model is known to exhibit a topological phase transition, characterized by the absence of a local order parameter \cite{RevModPhys.82.3045, Alicea_2012}.  It is defined by a one-dimensional chain of $N$ spinless fermions, whose creation and annihilation operators obey anticommutation rules: $\{a_i,a_j^\dag\}=\delta_{ij}$ and $\{a_i^\dag,a_j^\dag\}=\{a_i,a_j\}=0$, where $i,j$ are site indices. Its Hamiltonian is given by \cite{Kitaev_2001}
%\begin{eqnarray} \label{model}
% H= \sum_j \Bigl[ -wa_j^\dagger a_{j+1}+\Delta a_j a_{j+1} + h.c.\Bigr]
% -\mu(a_j^\dagger a_j-{\textstyle\frac{1}{2}})
%\end{eqnarray}
\begin{eqnarray} \label{model}
H&=& \sum_j \Bigl[ -w(a_j^\dagger a_{j+1}+a_{j+1}^\dagger a_j) \nonumber
\\&-&\mu(a_j^\dagger a_j-{\textstyle\frac{1}{2}})
+\Delta( a_ja_{j+1}+ a_{j+1}^\dagger a_j^\dagger) \Bigr],
\end{eqnarray}
where $\mu$ is the chemical potential, $w$ is the hopping amplitude, and $\Delta$ (assumed to be real) is the p-wave superconducting pairing constant.  

%The Kitaev model is of great importance in topological superconductivity because of its ability to accommodate Majorana zero modes at its edges \cite{Kitaev_2001,aguado2017majorana}. Although the Majorana representation provides \glb{deep insights into the model's underlying physics}, its inclusion is beyond the scope of our work. Our focus is on the application of complex network analysis, which requires only the calculation of the ground state density matrix.
The Kitaev model plays a pivotal role in topological superconductivity due to its ability to host Majorana zero modes at the edges \cite{Kitaev_2001,aguado2017majorana}. While the Majorana representation offers deep insights into the model's underlying physics, its detailed treatment lies beyond the scope of this work. Instead, we focus on applying complex network analysis, which relies solely on the ground-state density matrix.
In this work, we focus on the case of finite size, where precursors of quantum criticality can be identified. To provide a comprehensive analysis, we study both periodic and open boundary conditions, aiming to capture the different features and behaviors that arise in each scenario.

 A key symmetry property of the Kitaev model is revealed through the parity operator $P=\prod_{i=1}^N (1-2a_{j}^{\dagger }a_{j})$, which has eigenvalues $\pm 1$ and commutes with the Hamiltonian $[H,P]=0$. 
 %A fundamental property of the Kitaev model is its symmetry: defining the parity operator $P=\prod_{i=1}^N (2a_{j}^{\dagger }a_{j}-1)$, whose eigenvalues are $\pm 1$, we have $[H,P]=0$. 
 This commutation relation implies that, in the absence of degeneracy, every eigenstate of $H$ has definite parity - it is simultaneously an eigenstate of $P$ with eigenvalue determined by the fermion number module $2$.
 %This implies that, in the absence of degeneracy, any eigenvector of $H$ belongs to a symmetry sector; that is, it is also an eigenvector of $P$ with a well-defined parity (as determined by the number of fermions).
 
For periodic boundary conditions, the Kitaev model is diagonalized via successive Fourier and Bogoliubov transformations, giving the quasiparticle spectrum
%written in its diagonal form as
\begin{equation}
    H=\sum_{k=0}^{N-1} \Lambda_k \left(b_k^\dag b_k-\frac{1}{2}\right),
\end{equation}
where $\Lambda_k=\sqrt{\epsilon_k^2+|\Delta_k|^2}$, with $\epsilon_k=-\mu-2 w \cos (2 \pi k/N)$ and $\Delta_k=2 i \Delta \sin(2 \pi k/N)$, and $b_k$ ($b_k^\dag$) are new fermionic annihilation (creation) operators. %Finally, the (nonlocal in the lattice space) fermionic operators $a_k^\dag$ and $ a_k$ can be obtained from $a_j^\dag$ and $ a_j$ by means of a Fourier transform followed by a Boguliubov transformation. 
The parity of the ground state  %\sout{is \glb{not trivial to compute. It is} related to the sign of the Pfaffian of the Hamiltonian in an antisymmetric basis and }
depends critically on the energies of the modes $k=\{0,N/2\}$ \cite{Kitaev_2001}. As a result, the ground state has odd parity in the range $-2 w<\mu<2 w$ and even elsewhere. In the thermodynamic limit, this change in symmetry at $\mu=\pm2w$ gives rise to the  phase transition between the trivial and topological phases. Further details can be found in Appendix \ref{appendix:trivial_and_topological}.
For open boundary conditions, the solution is less straightforward and can be found in Refs. \cite{PhysRevB.90.245435,leumer2020exact}. Of particular relevance to our analysis is the existence of $N$ distinct values for the chemical potential, 
$\mu_n=2\sqrt{w^2-\Delta^2 }\cos[\pi n/(N+1)]$
with $n=1,\dots,N$, which correspond to Majorana zero-mode energies. Around these values, the ground state undergoes parity switching \cite{hegde2015quench}. Within the range $0<\mu <2w$, there are $N/2$  parity transitions for even $N$ and $(N-1)/2$ transitions when $N$ is odd.

While in the thermodynamic limit, the Kitaev model is known to be formally (although not physically) equivalent to the  Ising spin model \cite{greiter20141d},  a crucial distinction arises for finite systems. The Jordan-Wigner transformation —which establishes a spin-fermion mapping— is only exact within a single parity sector. We will examine this important limitation in detail later in this work.

\begin{figure}[ht]
    \centering
    \includegraphics[width=0.47\textwidth]{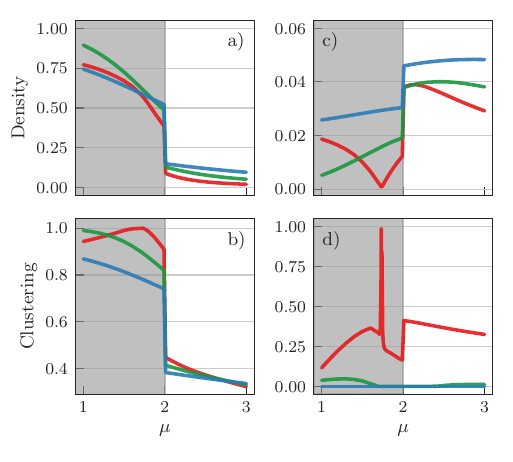}
    \caption{
    Density (a,c) and clustering (b,d) for different correlation networks of $N=14$ sites with $\Delta=0.5, 1.0, 2.0$ depicted in red, green, and blue, respectively. Grey (white) region represents the topological (trivial) phase. In panels (a-b), the quantum mutual information is examined as a function of the chemical potential $\mu$. Both the density (a) and the clustering (b) detect the topological phase transition at $\mu_c \equiv 2w = 2$. In panels (c-d), the entanglement network is investigated using the concurrence. Similar to the mutual information case, the density (c) accurately captures the phase transition. However, the clustering network (d) is not sensitive enough for $\Delta \geq 1$. Both for the mutual information and entanglement, a point with $C=1$ is also observed.
    %\sout{ Unexpectedly, a new critical point with clustering ($C=1$) is found for $\Delta=0.5$, indicating a homogeneous network structure. This implies that, in a system with periodic boundary conditions, the concurrence remains constant and does not decay with the distance.}
    }
    %In addition, the density \textbf{(c)} indicates that the concurrence is low at $\mu_c$. Although, for $\Delta=0.5$ the extreme points in \textbf{(c-d)} appear similiar, it might not hold for other values of $\Delta$.}}
    \label{fig:networks with PBC}
\end{figure}

%----------------------------------------------%
\section{Methods}  \label{sec:methods}
%----------------------------------------------%

The correlation-based network approach begins by defining an appropriate bipartite correlation measure $\mathcal{M}$ between lattice sites \cite{carr2017}. In our analysis, we employ two fundamental measures: (i) the quantum mutual information $\mathcal{I}$, which captures total correlations, and (ii) the entanglement $\mathcal{E}$, quantified through concurrence \cite{PhysRevLett.80.2245}.
The behavior of the coherence $\mathcal{C}$ is also addressed in Appendix \ref{appendix:coherence}.

The calculation proceeds as follows: starting from the ground state $\ket{G}$,  we compute the reduced density matrix $\rho_{ij}$ for each site pair $(i,j)$
%The procedure begins with the ground state $\ket{G}$ of our system. For every pair of sites $i$ and $j$, we compute the reduced density matrix $\rho_{ij}$ 
and use the corresponding correlation measure $\mathcal{M}$ to construct a bidirectional complex network, which consists of $N$ nodes and $N(N-1)/2$ edges with strength $e_{i,j}\equiv\mathcal{M}_{i,j}$. All numerical simulations and network analyses were performed using the open-source code available at \cite{llodra2025kitaev}. Although our analysis focuses on $\ket{G}$, the framework is easily extended to finite temperatures via the Gibbs density matrix \cite{sundar2018complex}.

%The network complexity can be characterized through multiple topological indicators, each capturing distinct structural properties. While our investigation considered several such measures, we focus here on two particularly informative metrics: (i) the network density, quantifying connection prevalence, and (ii) the clustering coefficient, measuring local interconnectivity.
The complexity of such networks can be characterized using a variety of indicators, each capturing distinct structural properties. While our investigation considered several such measures, we focus here on two particularly informative metrics: the density and the clustering coefficient. The density of a node $i$ ($d_{i}$) quantifies its relative importance within the network. It is defined as the ratio between the total weights of links connected to $i$ and the maximum number of connections it could have:
\begin{equation}
     d_{i}=\frac{\sum_{j\neq i}e_{ij}}{N-1}.
\end{equation}
The clustering coefficient ($C$) measures the tendency of nodes to form tightly connected groups or clusters. For a weighted network, it is defined as
\begin{equation}
    C=\frac{\sum_{i \neq j \neq k} e_{ij} e_{jk} e_{ki}}{\sum_k \sum_{i \neq j \neq k} e_{ik} e_{jk}},
\end{equation}
and quantifies the fraction of weighted closed triangles relative to the total number of weighted triplets in the network \footnote{Intuitively, in an unweighted network, this corresponds to the ratio of closed triplets (three nodes forming a triangle) to open triplets (three nodes where only two edges exist, forming a V shape). While our networks are weighted, this interpretation provides useful intuition}. Increased clustering coefficient implies stronger local connectivity and a greater likelihood of nodes forming interconnected groups.
%A higher clustering coefficient indicates a stronger tendency for nodes to form interconnected structures.

%---------------------------------------%
%\subsection{Periodic boundary conditions}
\section{Results} \label{sec:results}
%---------------------------------------%

Let us start our analysis with the case of a closed chain. Due to the translational invariant of the system, all sites are equivalent, leading to regular networks. In Fig. \ref{fig:networks with PBC}a-\ref{fig:networks with PBC}b we consider a network built using the mutual information $\mathcal{I}$ and look at the behavior of density ($d_i$) (which is actually independent of $i$) and the clustering coefficient ($C$). These are plotted as functions of the chemical potential, for a fixed value of the hopping $w=1$, which we will use to scale all other energy values, and for three different values of $\Delta$. Both the $d_i$ (upper panels) and $C$ (lower panels) exhibit a sharp discontinuity around $\mu=\mu_c\equiv 2$, where the ground state symmetry changes. Notably, the clustering also displays a pronounced maximum $C=1$ for a value of $\mu^*<\mu_c$, see Fig. \ref{fig:networks with PBC}b. This feature is only observed for the case $\Delta=0.5$.  We will examine this distinctive behavior in detail in Section \ref{sec:analytical}.

\begin{figure}[b]
    \centering
    \includegraphics[width=0.48\textwidth]{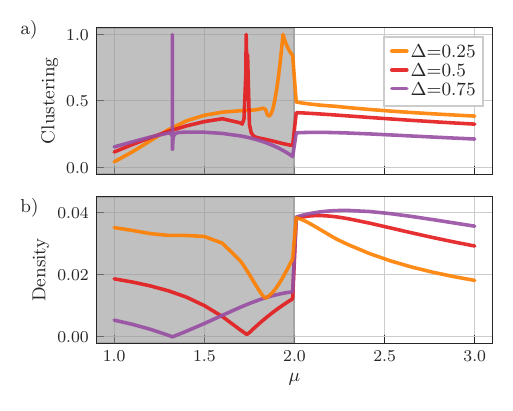}
    \caption{Detection of $C=1$ for a concurrence network with $\Delta=0.25, 0.5, 0.75$. In panel (a), the sharp peak observed numerically closely follows the analytical prediction, $\mu^*=2\sqrt{w^2-\Delta^2}$ derived in section \ref{sec:analytical}. Although the maximum peak in (a) aligns closely with the minimum peak in (b), both peaks have different values of $\mu$. The resolution near the peaks is $10^{-3}$ and $10^{-4}$ for the purple line.}
    %Concurrence networks with identical weights.
    %The same correlation network as in Fig. \ref{fig:networks with PBC}d  %also denotes the presence of maximal clustering 
    %for  \glb{comment density}. %Clearly, this hidden feature ($C=1$)in the Kitaev depends on $\Delta$. As explained in Section \ref{sec:section 5}, this entanglement structure emerge at $\mu^* \equiv 2\sqrt{w^2-\Delta^2}$.
    %In the periodic case, we establish a connection between the XY Ising model and this feature ($C=1$) from the Kitaev model, which predict the existence of an homogenuos entanglement structure at $\mu^* \equiv 2\sqrt{^2-\Delta^2}$ (see Section \ref{sec:section 5} for more details).} }
    \label{fig:clustering=1}
\end{figure}

In Fig. \ref{fig:networks with PBC}c-\ref{fig:networks with PBC}d we examine a network built using the entanglement (measured by the concurrence) between all pairs of sites. The concurrence network can also capture the symmetry change around $\mu=\mu_c$, though only for $\Delta=0.5$ in the case of clustering. However, in this case, the clustering is much more pronounced around $\mu^*$. %The density shows a sharp minimum for a different value of $\mu$.  
Notice that the sharp maximum in the clustering is not related to the minimum denoted by the density. This difference is more evident for $\Delta=0.25$ in Fig \ref{fig:clustering=1}.

To better understand the meaning of $\mu=\mu^*$, in Fig. \ref{fig:clustering=1} we extend the analysis from Fig. \ref{fig:networks with PBC}c-\ref{fig:networks with PBC}d to additional values of $\Delta<1$. In every case, we find a distinct $\mu^*(\Delta)$ where the clustering reaches its maximum ($C=1$). Therefore, this reveals that within the ``topological'' phase there exists a special value of the chemical potential {$\mu^*$} such that the network becomes fully connected, with all nodes linked by edges of equal weight. Motivated by this numerical evidence, we found an analytical explanation for the existence of such a singularity by comparing the differences and similarities between the Kitaev model and the XY spin chain. We anticipate that this singular behavior is closely related to the phenomenon of ground state factorization in spin chains.   

%---------------------------------------%
%\subsection{Open boundary conditions}
%\label{sec:open-boundaries}
%---------------------------------------%

%In finite-size systems with open boundaries, as discussed previously, 
Now, let us consider a finite-sized system with open boundary conditions. As discussed previously,
the Kitaev model's ground state displays characteristic parity sector jumps, indicative of gap closure and Majorana zero mode formation. In this case, the system ceases to be translationally invariant, resulting in position-dependent site densities $d_i$.

To illustrate the implications of these symmetry transitions and the emergence of Majorana zero modes, we analyze the behavior of the clustering coefficient and density in the concurrence network in Fig. \ref{fig:open_boundaries}. For a discussion of the mutual information network, see Appendix \ref{appendix::mutual information}. The top two panels show the clustering coefficient and density for different values of $\Delta$, while the lower panel shows the clustering coefficient variation for different system sizes. In all cases, we observe sharp discontinuities precisely at the chemical potential values $\mu_n$, where Majorana zero modes are expected to appear. %These discontinuities signify the vanishing of the energy gap in the ground state symmetry and consequently the appearance of Majorana zero modes. 
Also, note that the singular behavior of the clustering $C=1$ observed for periodic boundary conditions has no analog here.

\begin{figure}[t]
    \centering
    \includegraphics[width=0.47\textwidth]{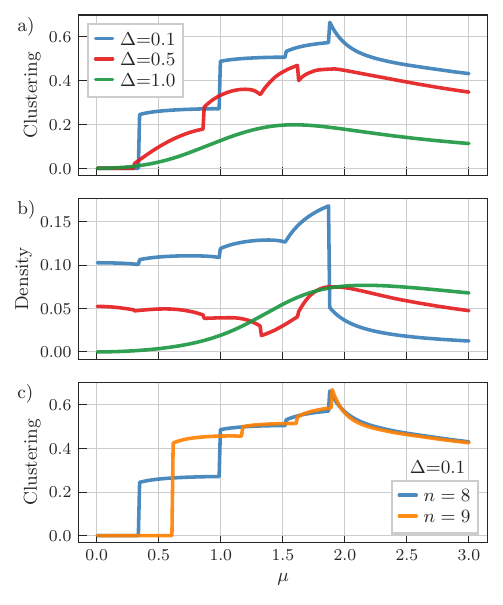}
    \caption{
    Zero-energy modes in the open chain. The clustering (a) and density (b) metrics of the concurrence network capture the presence of Majorana zero-energy modes in the Kitaev chain with open boundary conditions. (a-b) Four zero-energy modes are detected for $\Delta=0.1, 0.5$ with $N=8$ sites, appearing at positive $\mu$-values. The remaining four zero-energy modes are symmetrically located at negative $\mu$-values. (b) Clustering coefficient with $N=8$ and $N=9$ fermion sites.}
    %\glb{\sout g    {In addition,  $\Delta=1$ confirms that there are not zero energy modes when $w^2 \geq \Delta^2$.}}}
    \label{fig:open_boundaries}
\end{figure}

%---------------------------------------%
%\glb{\section{Analytical proof of the hidden features}}\label{sec:complete}
\section{Analytical explanation for the complete graph}
\label{sec:analytical}
%---------------------------------------%

As we have seen, building correlation networks from the Kitaev chain successfully captures all symmetry changes in the ground state. For the periodic chain, this approach identifies the critical chemical potential that anticipates the topological phase transition in the thermodynamic limit. Under open boundary conditions, it also reveals the complete set of Majorana zero modes. 

Beyond these well-established results, our complex network approach uncovers a previously unnoticed singularity at a special value of the chemical potential, $\mu^*$, where the network becomes fully regular, i.e., all nodes are connected and all weights are equal. 
This means that the density matrix of the ground state $\rho$ is invariant under any site permutation. Then, the reduced density matrix for any pair  of fermions, $\rho_{i,j}={\rm Tr}_{\bar{i},\bar{j}} [\rho]$, is independent of $i$ and $j$. Interestingly, standard indicators such as ground state fidelity \cite{PhysRevE.74.031123,PhysRevLett.99.100603} (see  Fig. \ref{fig:topological transition} in Appendix \ref{appendix:trivial_and_topological}), provide no evidence for this singular behavior, which to our knowledge has not been reported before. The numerical values we found for $\mu^*$ suggest an analytical explanation can be found by looking at the analogies and differences between the Kitaev and XY spin models. We will show that the complete graph structure is the counterpart of the well-known ground state factorization in spin chains. As a side note, it is worth mentioning that the ground state factorization has also been observed in the Kitaev model under open boundary conditions with carefully tuned local chemical potentials \cite{PhysRevB.92.115137}. 

%Let us sketch the derivation for the emergence of XXX here, leaving the technical details in the appendix. 
First of all, let us recall that the cyclic XY model describing $N$ spins-$1/2$ is commonly defined as    
\begin{equation}\label{eq:xy}
% H_{XY}=-\frac{1}{2}\sum_{i=1}^N \left[( J+\gamma)\sigma_i^x\sigma_{i+1}^x + ( J-\gamma)\sigma_i^y\sigma_{i+1}^y+h \sigma_i^z\right]\label{xy}
H_{XY}=-\sum_{i=1}^N \left(J_x\sigma_i^x\sigma_{i+1}^x + J_y \sigma_i^y\sigma_{i+1}^y\right)-\frac{h}{2}\sum_{i=1}^N \sigma_i^z
%\label{hb}
\end{equation}
% \begin{equation}
% H_{XY}=-\sum_{i=1}^N \left[\left( \frac{J+\gamma}{2}\right)\sigma_i^x\sigma_{i+1}^x + \left( \frac{J-\gamma}{2}\right)\sigma_i^y\sigma_{i+1}^y\right]-\frac{h}{2}\sum_{i=1}^N \sigma_i^z\label{xy}
% \end{equation}
where periodic boundary conditions are ensured defining $\sigma^\alpha_{N+1}\equiv \sigma^\alpha_{1}$, with $\alpha=x,y,z$. As in the case of the Kitaev model, the XY model Hamiltonian commutes with the parity operator, now defined as $P_{\rm XY}=\prod_i \sigma_i^z$, whose eigenvalues are $p=\pm 1$. This implies that all eigenvectors can be classified as either even ($p=+1$) or odd ($p=-1)$. This system shows the peculiar phenomenon of ground state factorization (GSF) \cite{KURMANN1982235,PhysRevLett.94.147208,Giampaolo}. In short, assuming for simplicity $J\equiv J_x+J_y=1$ and defining $\gamma\equiv J_x-J_y$, for $h=2\sqrt{J^2-\gamma^2}$ there are two degenerate ground states that can be written as 
\begin{equation}
    |\Psi_F^{\pm}\rangle=\otimes_{i=1}^N(\cos\theta |\uparrow\rangle\pm \sin\theta |\downarrow\rangle),
\end{equation}
where $\theta=\arctan\sqrt{\tan \phi}$, with $\phi=(\arcsin \gamma)/2$. As discussed in Refs. \cite{rossignoli,giorgi2009ground}, these two states break the Hamiltonian parity symmetry and can be expressed as linear combinations of the even- and odd-parity eigenstates \cite{rossignoli}: 
\begin{equation}
    |\Psi_F^{\pm}\rangle=(|\Psi_F^{\rm even}\rangle\pm |\Psi_F^{\rm odd}\rangle)/\mathcal{N},
\end{equation}
being  $\mathcal{N}$ a normalization factor.

To establish a comparison between the Kitaev model and the XY chain, we will make use of the Jordan-Wigner transformation \cite{lieb1961two}, which connects their respective degrees of freedom. 
 First, let us choose $J=w$, $\gamma=\Delta$, and $h=\mu$.  
%In the thermodynamic limit the two models can be mapped into each other through the Jordan-Wigner transformation []. Consequently,  they share the same set of eigenvalues and each eigenvector of one of the systems can be mapped onto a corresponding one of the other system. 
For the sake of simplicity in this derivation, we will also assume an even number of sites $N$ (this does not change any of the results).
In the case of finite-size models, as described in Appendix \ref{appendix:map_xy_kitaev}, the two models can be mapped into each other only for the part of the eigenvectors and eigenvalues belonging to the odd symmetry sector, while such a mapping %the mapping between the two models holds exactly for the part of the eigenvectors and eigenvalues belonging to the odd symmetry sector, while it 
does not hold in the even symmetry sector. Noting that at the point 
\begin{equation}
    \mu^*=2\sqrt{w^2-\Delta^2},
\end{equation} 
which corresponds to GSF for the spin chain, the lower energy state of the Kitaev model has odd parity %(we are within the range $-2 w<\mu<2 w$), 
and then can be analytically determined by applying the Jordan-Wigner mapping to  $|\Psi_F^{\rm odd}\rangle$. 
%This implies that at the point $\mu=2\sqrt{w-\Delta^2}$ (corresponding to GSF for the spin chain), the lower energy state of the Kitaev model, which has odd parity in the range $-2 w<\mu<2 w$,  can be analytically determined by applying the Jordan-Wigner mapping to  $|\Psi_F^{\rm odd}\rangle$. 
According to Ref. \cite{rossignoli}, such a state can be written as 
% \begin{equation}
% |\Psi_F^{\rm odd}\rangle=\sum_{k \;{\rm odd} } \frac{\sqrt{2}\sin^k\theta \cos^{N-k}\theta}{k!\sqrt{1-\cos^N 2\theta}} S_-^k| \uparrow,\uparrow,\cdots, \uparrow\rangle,
% \end{equation}
\begin{equation}
|\Psi_F^{\rm odd}\rangle=\sum_{k \,{\rm odd} }  f_{N,k}(\theta) (S^-)^k| \uparrow,\uparrow,\cdots, \uparrow\rangle,
\end{equation}
where  $S^-=\sum_{j=1}^N\sigma^-_j$ is the collective spin operator and where $f_{N,k}^{(\theta)}=\frac{\sqrt{2}\sin^k\theta \cos^{N-k}\theta}{k!\sqrt{1-\cos^N 2\theta}}$.  
Due to the form of the weights $f_{N,k}^{(\theta)}$, this state is invariant under any permutation operator and exhibits uniform nonzero entanglement between any pair of spins \cite{rossignoli}.
%As a consequence, the ground state of the Kitaev model inherits the same property. Indeed, 
By applying the Jordan-Wigner transformation to $S^-$, the fermionic ground state is
\begin{equation}
|\Psi\rangle=\sum_{k \;{\rm odd} } f_{N,k}^{(\theta)}\left[ \sum_{i=1}^N \prod_{j<i}\left( 2a_{j}^{\dagger }a_{j}-1\right) a_{i}\right]^k |1,1,\cdots,1\rangle,
\end{equation}
which is also invariant under site permutation because of $f_{N,k}^{(\theta)}$. This guarantees that by performing the partial trace over all fermions except $i$ and $j$ we obtain exactly the same reduced density matrix, so that the classical network associated with such a state has maximum clustering, as we have explained before.

%that the lower-energy  state of the odd sector, which turns out to be the ground state and can be calculated exactly starting from Eq. (...), has long-range, uniform correlations, and thus gives rise to the regular network detected in Figs...  

%---------------------------------------%
\section{Conclusions}\label{sec:conclusions}
%---------------------------------------%
Our study effectively implemented the complex network approach to examine the ground state of the Kitaev model. When periodic boundary conditions were considered, the translationally invariant network displayed its capability to discern the critical point, indicating a symmetry change in the ground state that predicts the topological phase transition in the thermodynamic limit.

Furthermore, our approach has proven to be effective in detecting all Majorana zero modes present in the finite chain under open boundary conditions. Although some correlation networks were not equally successful in detecting all transitions (especially, entanglement showed greater sensitivity and could disappear over a larger region around critical points), our findings indicate that the correlation network method has enough efficacy to identify phase transitions beyond the conventional Landau paradigm, as demonstrated by the Kitaev model.

%In conclusion, we have applied the complex network approach to the ground state of the Kitaev model. In the case of periodic boundary conditions, the (translationally invariant) network can detect the presence of a critical point, which corresponds to a symmetry switch in the ground state and anticipates the topological phase transition in the thermodynamic limit. In the case of open boundary conditions, we were able to detect all Majorana zero modes present in the finite chain. While not all correlation networks were able to detect all transitions (entanglement turns out to be more fragile, as it can vanish in a wider region around critical points), in general, the technique seems to be powerful enough to detect phase transitions beyond the Landau paradigm, as in the case of the Kitaev model. 

More interestingly, our study using the complex network approach uncovered a remarkable discovery —the existence of a long-range, highly symmetric ground state in the Kitaev model that remains invariant under any possible site swap. This intriguing property was revealed by a complete homogenous graph within the network representation.
Through  analytical arguments, we established a connection between this unique behavior of the Kitaev model and the phenomenon of ground state factorization observed in the Ising spin chain. Our findings demonstrate that the symmetry inherent in the Kitaev model serves as a direct analog to ground state factorization in the Ising chain, unveiling a deeper conceptual connection between these two systems. This insight not only enhances our understanding of the Kitaev model but also provides a unifying perspective on the symmetries and factorization properties in  broader classes of quantum spin systems.

%Through analytical arguments, we established a connection between this singular behavior of the Kitaev model and the ground state factorization observed in the Ising spin chain. Remarkably, this finding suggests that this symmetry in the Kitaev model can be seen as the counterpart of the ground state factorization found in the Ising spin chain.

The importance of this outcome is not only in its  specific implications for the Kitaev model but also in its broader impact on the application of the complex network approach to quantum systems. 
Although the Kitaev model has been thoroughly studied, the complex network analysis revealed a previously undiscovered structure characterized by its simplicity, which had eluded all prior analytical methods. 
This finding underscores the power of the complex network approach in revealing latent properties of quantum systems, providing new perspectives, and exposing intricate behaviors that might have been hidden by standard analytical techniques.

\section*{Data Availability Statement}
All data and code used in this study are openly available in the repository \url{https://github.com/gllodra12/KitaevNetworkAnalysis}.

\appendix
\renewcommand{\thefigure}{A\arabic{figure}}
\setcounter{figure}{0}
\section{Trivial and topological phase}
\label{appendix:trivial_and_topological}

\begin{figure}
    \centering
    \includegraphics[width=0.47\textwidth]{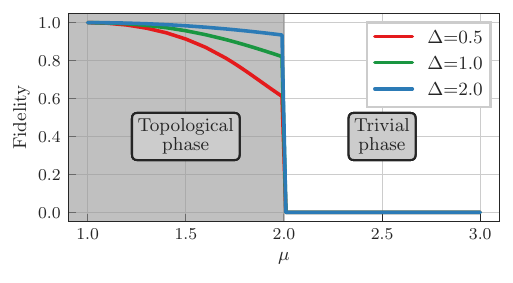}
    \caption{
    %\glb{\bf{Topological phase transition.}} 
    The Kitaev chain with periodic boundary conditions has a ground state with an odd (topological phase) or even (trivial phase) parity. The fidelity, $ \langle %\psi_{gs}
    G(\mu=1)|G(\mu) \rangle$, quantifies the overlap between the ground state at $\mu=1$ and the ground states for $\mu \in [1, 3]$. Near the quantum critical point $\mu_t = 2 w$ the ground state jumps from one symmetry sector to another. This transition is marked by a sharp drop in the fidelity, which approaches zero at $\mu=2$ (note that $w=1$ and $N=14$ unless specified otherwise).} 
    \label{fig:topological transition}
\end{figure}

In the main text, we introduced the Kitaev model, whose ground state exhibits a change in parity with respect to the operator $P=\prod_{i=1}^N (1-2a_{j}^{\dagger }a_{j})$. Specifically, the ground state has odd parity when the chemical potential satisfies $\mu \in [-2w, 2w]$ and an even parity outside this interval. To illustrate this symmetry change more explicitly, Fig. \ref{fig:topological transition} represents the fidelity $\mathcal{F}(\mu)=\langle G(\mu=1) | G(\mu)\rangle$ where $|G(\mu=1)\rangle$ is the ground state computed at $\mu=1$, and $|G(\mu)\rangle$ is the ground state for a variable $\mu$. The fidelity $\mathcal{F}(\mu)$ quantifies the overlap between these two states. As $\mu$ approaches the critical value, $\mathcal{F}(\mu)$ sharply drops to zero. This behavior indicates that the ground states on either side of the transition belong to orthogonal symmetry sectors. 
%The vanishing overlap %at the critical point  represents a precursor of the topological phase transition %, reinforcing the idea that the symmetry of the ground state changes precisely  at $\mu=\pm 2w$. 
The vanishing overlap between different ground states
represents a precursor of the topological phase transition 
 at $\mu=\pm 2w$. A similar result was also shown by Ref. \cite{chou2014network} using the pairing amplitude over the ground state wave function.

\renewcommand{\thefigure}{B\arabic{figure}}
\setcounter{figure}{0}
\section{Coherence correlation network}
\label{appendix:coherence}

\begin{figure}
    \centering
    \includegraphics[width=0.48\textwidth]{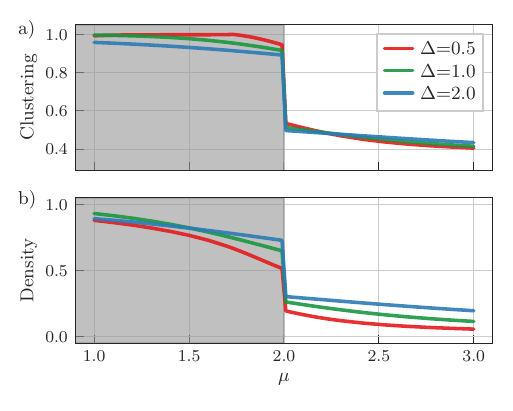}
    \caption{Clustering (a) and density (b) in a coherence-based correlation network for the periodic Kitaev chain, shown as function of the chemical potential $\mu$. Each curve corresponds to a different pairing amplitude $\Delta$, with $\Delta=0.5$ (red), $\Delta=1.0$ (green), and $\Delta=2.0$ (blue). Both the clustering and the density detect the topological phase transition at $\mu=2w$. While the additional critical point at $\mu^*=\sqrt{3}$ is captured by the clustering coefficient, its signature is subtler here than in the mutual information or concurrence-based networks.}
    \label{fig:coherence}
\end{figure}

In section \ref{sec:results}, we examined the symmetry transition of the ground state using networks constructed from quantum mutual information and concurrence. Here, we broaden our analysis by introducing an alternative network framework that employs quantum coherence measures (see Fig. \ref{fig:coherence}). These results reinforce our earlier findings: the topological phase transition is clearly detected at $\mu=2w$, and the critical point $\mu^*=2\sqrt{w^2-\Delta^2}$ is identified in Fig. \ref{fig:coherence}a. Although the coherence-based network reveals the critical point $\mu^*$ with a subtler signature compared to the networks based on mutual information and concurrence, it nevertheless confirms the presence of a complete homogeneous network since the clustering reaches the value one.% at $\mu=\sqrt{3}$.

 \renewcommand{\thefigure}{C\arabic{figure}}
 \setcounter{figure}{0}
 \section{Open boundaries with mutual information}
 \label{appendix::mutual information}
\begin{figure}
    \centering
    \includegraphics[width=0.48\textwidth]{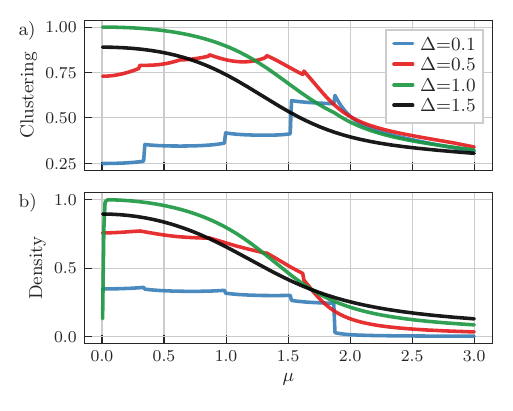}
    \caption{Clustering (a) and density (b) in a mutual information-based correlation network for the open Kitaev chain, shown as a function of the chemical potential $\mu$. The different curves correspond to $\Delta=0.1$ (blue), $\Delta=0.5$ (red), $\Delta=1.0$ (green), $\Delta=1.5$ (black). Both the clustering and density metrics reveal how the underlying correlation structure changes with $\mu$,  highlighting the impact of increasing superconducting pairing $\Delta$ on the network's topology and signaling symmetry transitions for the different values of $\mu_n=2\sqrt{w^2-\Delta^2 }\cos[\pi n/(N+1)]$.} 
    \label{fig:appendix mutual}
\end{figure}

In this section, we explore how the zero-energy modes present in the open Kitaev chain can be detected using the mutual information correlation network. For a more complete picture, we present the behavior of the clustering and density in Fig. \ref{fig:appendix mutual}. As discussed in the main text, the alternating behavior observed in Fig. \ref{fig:appendix mutual} occurs at specific values of the chemical potential, $\mu_n$, which are indicative of transitions in the parity of the ground state. These transitions are directly linked to the emergence of Majorana zero modes, which alter the symmetry properties of the system. Notably, these oscillations become apparent only when the condition $|\Delta| < |w|$ is met, ensuring that the pairing strength is sufficiently weak relative to the hopping amplitude.

It is important to highlight that while these oscillations in the mutual information network do signal the underlying topological transitions, their amplitude is generally less pronounced compared to the features observed in the concurrence-based network (see Fig. 3 in the main text). In particular, as the superconducting pairing parameter $\Delta$ approaches unity, the oscillations in the mutual information network become increasingly subdued. This difference in visibility may reflect the distinct ways in which mutual information and concurrence capture quantum correlations in the system.

\renewcommand{\thefigure}{D\arabic{figure}}
\setcounter{figure}{0}
\section{Mapping the XY model into Kitaev}
\label{appendix:map_xy_kitaev}

Even though the XY and Kitaev models describe distinct physical systems, they are mathematically connected through a mapping known as the Jordan-Wigner transformation \cite{lieb1961two}.
We anticipate here that the mapping is exact in the thermodynamic limit, while for finite systems it is only valid for a symmetry sector. The Jordan-Wigner transformation is defined as 
\begin{eqnarray}
\sigma _{i}^{z}&=&1- 2a_i^\dagger a_i,\\
\sigma_{i}^{-}&=&\prod_{j<i}\left( 1-2a_j^\dagger a_j\right) a_{i},\\
\sigma _{i}^{+}&=&\prod_{j<i}\left( 1-2a_j^\dagger a_j\right)a_{i}^{\dagger },
\end{eqnarray}
where  the string operators $\prod_{j<i}\left( 1-2n_{j}\right)$ have the function to convert commutators into anticommutators.
%\begin{eqnarray}
%\sigma _{i}^{z}&=&2a_{i}^{\dagger }a_{i}-1\\
%\sigma_{i}^{-}&=&\prod_{j<i}\left( 1-2a_{j}^{\dagger }a_{j}\right) a_{i}\\
%\sigma _{i}^{+}&=&\prod_{j<i}\left( 1-2a_{j}^{\dagger }a_{j}\right)a_{i}^{\dagger }
%\end{eqnarray}
Applying such a transformation to the cyclic XY model (see Eq. (\ref{eq:xy}) in the main text), we get the new  Hamiltonian described in terms of fermionic operators $\tilde H_{XY}=\tilde H_{XY}^{(0)}+\tilde H_{XY}^{(1)}$, where
\begin{eqnarray}
\tilde H_{XY}^{(0)}&=&  -J\sum_{i=1}^{N-1} (a_j^\dagger a_{j+1}+a_{j+1}^\dagger a_j)
\\&-&h\sum_{i=1}^{N}(a_j^\dagger a_j-{\textstyle\frac{1}{2}})\nonumber
-\gamma\sum_{i=1}^{N-1}( a_ja_{j+1}+ a_{j+1}^\dagger a_j^\dagger) ,\nonumber\\
%\tilde H_{XY}^{(0)}&=&-h\sum_{i=1}^N (n_j%a_i^\dag a_{i}-\frac{1}{2})-\sum_{i=1}^{N-1} (J a_i^\dag a_{i+1} +\gamma a_i^\dag a^\dag_{i+1} +h.c.)\nonumber\\
\tilde H_{XY}^{(1)}&=&P  (Ja_N^\dag a_{1} +\gamma a_N^\dag a^\dag_{1} +h.c.),
\end{eqnarray}
%\begin{eqnarray}
%\tilde H_{XY}^{(0)}&=&h\sum_{i=1}^N \left(\frac{1}{2}-a_i^\dag a_{i}\right)-\sum_{i=1}^{N-1} %\tilde H_{XY}^{(1)}&=&P_K  (Ja_N^\dag a_{1} +\gamma a_N^\dag a^\dag_{1} +h.c.),
%\end{eqnarray}
with $J=J_x+J_y$ and $\gamma=J_x-J_y$ and where the parity (already defined in the main text for the Kitaev model) is $P=\prod_{i=1}^N (1-2a_j^\dagger a_{j})$.
% \begin{eqnarray}
% \tilde H_{XY}&=&-\sum_{i=1}^{N-1} [J(a_i^\dag a_{i+1} + a_{i+1}^\dag a_{i})+\gamma (a_i^\dag a^\dag_{i+1} + a_{i+1} a_{i})]\nonumber\\&-&h\sum_{i=1}^N (a_i^\dag a_{i}-1/2)+P_K  [J(a_N^\dag a_{1} + a_{1}^\dag a_{N})+\gamma (a_N^\dag a^\dag_{1} + a_{1} a_{N})]
% \end{eqnarray}
Then the odd part of the eigenvalues and eigenvectors of $\tilde H_{XY}$ is identical to the odd part of the eigenvalues and eigenvectors of the Kitaev Hamiltonian (provided that we identify $w=J,\; \Delta
=\gamma,\; \mu=h$). Nevertheless, this is not the case for the part of the chain corresponding to the even parity sector, which is mapped into an antiperiodic fermionic chain different from the Kitaev model. 
%This implies that (only) the odd part of the fermionic Hamiltonian is invariant under spatial translation, which was true for the whole Ising Hamiltonian.

%Using the same reasoning, if we started from the Kitaev model and tried to represent it in terms of spin operators, we would find an exact mapping between the odd parts of the two models, while the even part of the Kitaev chain would be mapped into an antiperiodic spin chain. Thus, in the finite-size limit, the Kitaev model loses the degeneracy corresponding to the ground state factorization of the Ising  model. Also note  that in the thermodynamic limit,  we can forget about boundary conditions, which makes the mapping exact. 

By applying a similar line of reasoning, if we were to consider the representation of the Kitaev model in terms of spin operators, we would find an exact mapping between the odd components of both models, while the even portion of the Kitaev chain would transform into an antiperiodic spin chain. This is the reason why, in the scenario of finite system sizes, the Kitaev model loses the degeneracy associated with the ground state factorization observed in the Ising model.
It is important to note that in the thermodynamic limit, the boundary conditions become irrelevant, allowing us to neglect their influence. This, in turn, makes the mapping between the two models exact.

\bibliography{references}% Produces the bibliography via BibTeX.

\end{document}